\newif\ifanonymoussubmission
\anonymoussubmissionfalse 
\ifanonymoussubmission
  \documentclass[sigplan,10pt,anonymous,review]{acmart}
\else
  \documentclass[sigplan,10pt]{acmart}
\fi
\setcopyright{none}
\acmConference[EuroSys '27]{European Conference on Computer Systems}{April 19--23, 2027}{Rabat, Morocco}
\acmYear{2027}
\acmDOI{}
\acmISBN{}
\usepackage{amsmath}
\usepackage{enumitem}
\usepackage{pifont}
\usepackage{listings}
\usepackage{tikz}
\usepackage{colortbl}
\usepackage{algorithm}
\usepackage{algorithmic}
\AtBeginDocument{\raggedbottom}
\AtBeginDocument{\mathchardef\originalmathcomma=\mathcode`\,\relax
  \mathcode`\,="8000\relax}
\begingroup
\catcode`\,=\active
\gdef,{\originalmathcomma\allowbreak}
\endgroup
\renewcommand{\small}{\fontsize{10}{12}\selectfont}
\renewcommand{\footnotesize}{\fontsize{10}{12}\selectfont}
\renewcommand{\scriptsize}{\fontsize{10}{12}\selectfont}

\newcommand{\cmark}{\ding{51}}
\newcommand{\xmark}{\ding{55}}
\newcommand{\papertitle}{Omni-Flow: A Unified Workflow Orchestration and Distributed KV Cache Sharing Framework for Multimodal Inference}
\hypersetup{pdfauthor={},pdfsubject={},pdfkeywords={}}

\begin{document}
\title[Omni-Flow]{\papertitle}
\ifanonymoussubmission\else
\author{Bin Xiao}
\authornote{Bin Xiao and Jingfu Dong contributed equally to this research.}
\correspondingauthor
\email{xiaobin14@meituan.com}
\affiliation{\institution{Meituan}\city{Beijing}\country{China}}

\author{Jingfu Dong}
\authornotemark[1]
\email{jfdong25@stu.pku.edu.cn}
\affiliation{\institution{Peking University}\city{Beijing}\country{China}}

\author{Changran Wang}
\email{wangchangran@meituan.com}
\affiliation{\institution{Meituan}\city{Beijing}\country{China}}

\author{Yitian Chen}
\email{chenyitian02@meituan.com}
\affiliation{\institution{Meituan}\city{Beijing}\country{China}}

\author{Xiaoyu Zhao}
\email{zhaoxiaoyu17@meituan.com}
\affiliation{\institution{Meituan}\city{Beijing}\country{China}}

\author{Yuqi Peng}
\email{pengyuqi@meituan.com}
\affiliation{\institution{Meituan}\city{Beijing}\country{China}}

\author{Jianping Lin}
\email{linjianping02@meituan.com}
\affiliation{\institution{Meituan}\city{Beijing}\country{China}}

\author{Yuchen Xie}
\email{xieyuchen@meituan.com}
\affiliation{\institution{Meituan}\city{Beijing}\country{China}}
\renewcommand{\shortauthors}{Xiao et al.}
\fi
\begin{abstract}
Multimodal models increasingly integrate heterogeneous components, from encoders and LLMs to diffusion models and media decoders. Serving these models efficiently requires flexible workflow orchestration, independent component scheduling, and cross-component state sharing. However, existing multimodal frameworks primarily organize execution as stage-level pipelines, while state ownership remains largely stage-local. We present Omni-Flow, a distributed serving framework built around three cooperating abstractions. Control Flow defines workflows through a Python DSL, organizing heterogeneous components and their dependencies into a unified dataflow graph. The runtime uses these dependencies to determine when each component can execute. Data Flow manages the placement, transfer, and lifetime of shared tensors and paged KV caches across roles, together with same-device weight sharing. Compute Flow matches multimodal conversation histories to reuse KV across turns and integrates framework-managed KV and model-specific sampling with SGLang's execution interfaces. Omni-Flow also enables cross-role prefix-KV reuse between compatible LLM and diffusion components. Together, these abstractions separate model-specific computation from workflow coordination and state management, allowing independently executing roles to share compatible resources through a common programming model. Experiments with Qwen3-Omni show that, across four benchmarks and multiple concurrency levels, Omni-Flow's aggregate job completion time (JCT) is 1.2\% and 4.9\% higher than vLLM-Omni's and SGLang-Omni's, respectively. In a 50-turn multimodal session, Omni-Flow avoids resubmitting accumulated inputs through server-side session state, reducing mean JCT by 7.5\% and 19.3\% relative to vLLM-Omni and SGLang-Omni, respectively.
\end{abstract}

\maketitle
\section{Introduction}
\label{sec:introduction}

\begin{figure}[t]
    \centering
    \includegraphics[width=\linewidth]{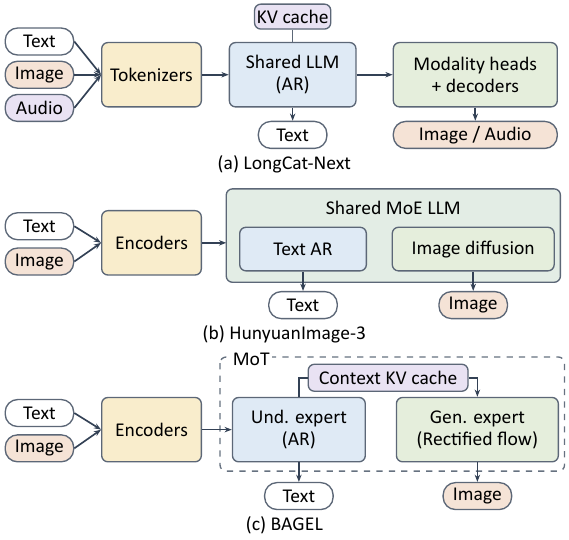}
    \caption{Unified multimodal architectures.}
    \label{fig:1}
\end{figure}

Multimodal models integrate modality-specific encoders, LLMs, diffusion models, and media decoders in a single inference process~\cite{llava,intern-vl,whisper,cosyvoice}. Serving them therefore requires coordinating a stateful workflow of heterogeneous components rather than merely running a single generation engine. Increasingly unified architectures, exemplified by BAGEL, HunyuanImage-3, and LongCat-Next, bring modality-specific or autoregressive and diffusion roles into a common model~\cite{bagel,hunyuanimage3,longcat-next} (Figure~\ref{fig:1}). Encoders process input segments, autoregressive engines advance token by token, and diffusion runtimes iteratively update latent states.

\begin{figure}[t]
    \centering
    \includegraphics[width=\linewidth]{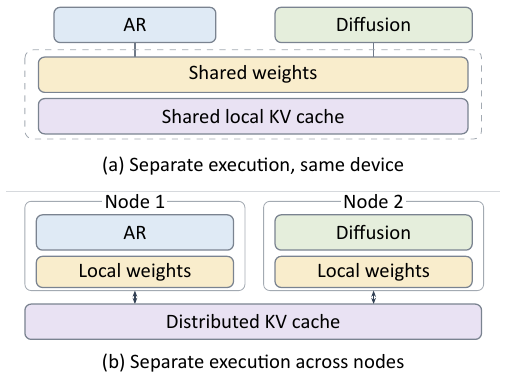}
    \caption{State sharing across deployment settings.}
    \label{fig:deployment-sharing}
\end{figure}

The central tension is between \emph{independent execution} and \emph{shared state}. Separating roles enables different batching, parallelism, and placement decisions, but reusable resources need not follow role boundaries. Context can outlive its producing invocation, and distinct roles may use compatible parameters. HunyuanImage-3, for example, combines autoregressive reasoning and diffusion-based image modeling in an LLM-based architecture~\cite{hunyuanimage3}. Reuse can span identical checkpoint parameters held by different role instances and previously computed context KV; it does not require all token types to use identical expert parameters. Separate runtimes can duplicate these parameters or require KV handoffs; a shared executor can couple their scheduling and placement. On the same device, independently executing roles can share compatible weights and local KV storage (Figure~\ref{fig:deployment-sharing}(a)); across nodes, weights remain local while reusable KV state can be managed through a distributed cache (Figure~\ref{fig:deployment-sharing}(b)). A serving framework must enable reuse without forcing these computations into one execution policy.

Existing engines provide paging, prefix reuse, and continuous batching~\cite{vllm,sglang,orca}; disaggregated serving and cache layers separate computation from portions of KV storage~\cite{distserve,mooncake,lmcache}; multimodal frameworks provide stage execution and transport~\cite{vllm-omni,sglang-omni-runtime}. Composing these facilities still requires contracts across engine boundaries for input completion, state ownership, and execution. Encoding those contracts separately for every model ties workflow semantics to model-specific runtime code. Three requirements guide our design.

\paragraph{Coordinate logical work.}
A streaming producer can yield repeatedly, and conditional branches can contribute different numbers of outputs. Receiving every required field need not mean that all contributions have arrived. Even complete inputs may need to wait for an earlier stateful invocation. The runtime must distinguish contribution grouping, completion, and execution order.

\paragraph{Manage state beyond role lifetimes.}
Intermediate tensors, KV prefixes, and checkpoint parameters have different reuse opportunities. Consumers must find compatible representations and keep storage valid independently of the producing invocation. This requires separating state identity, residency, and access lifetime while respecting model and parallel-layout constraints.

\paragraph{Separate execution reuse from scheduling.}
Autoregressive and diffusion roles can reuse compatible transformer infrastructure under different iteration policies. Multimodal KV reuse also needs replayable inputs: a cache identifier cannot recover missing modality embeddings. Execution interfaces must therefore expose storage below scheduling and retain the inputs needed for supported recomputation.

\begin{figure*}[t]
    \centering
    \includegraphics[width=\textwidth]{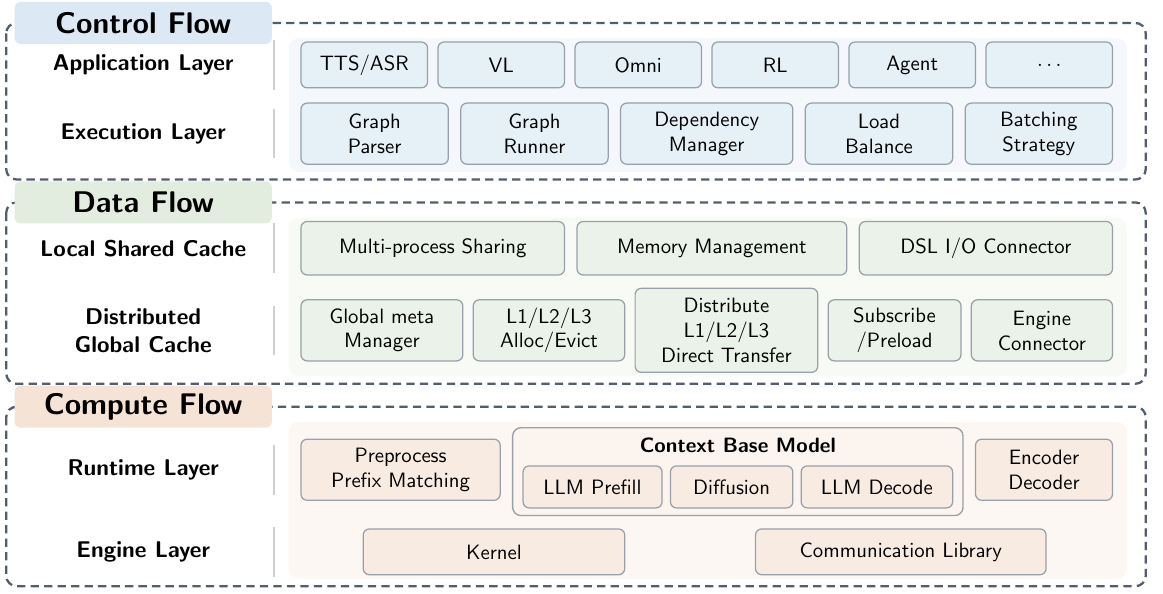}
    \caption{Omni-Flow architecture overview.}
    \label{fig:2}
\end{figure*}

We present \textsc{Omni-Flow}, a distributed serving framework with three cooperating abstractions (Figure~\ref{fig:2}). \textbf{Control Flow} represents workflows as declarative graphs, supports cycles, and determines which logical work may execute; \textbf{Data Flow} manages shared state and its lifetime; \textbf{Compute Flow} connects model execution to that state. This separation lets workflow readiness, cache availability, and engine scheduling evolve independently while preserving the contracts between them. Model integrations supply computational interfaces and workflow definitions; the framework supplies coordination and state management.

Our contributions are:
\begin{itemize}[itemsep=3pt,topsep=2pt,parsep=0pt,partopsep=0pt]
    \item \textbf{Causal workflow coordination.} A field-level graph combines inferred join scopes, causal leases, and explicit ordering to coordinate repeated yields, conditional paths, and request-dependent aggregation within validated graph structures.
    \item \textbf{State management across role boundaries.} A common data plane separates reusable-page identity from physical residency and role lifetime. Compatible-provider discovery and tiered KV management operate alongside intermediate-tensor access and same-device checkpoint sharing.
    \item \textbf{Execution adapters for heterogeneous reuse.} Adapters separate replayable multimodal inputs from materialized KV and connect framework-owned storage to SGLang. Compatible diffusion backbones reuse transformer execution and eligible weights and conditioning KV under separate diffusion scheduling.
\end{itemize}

The implementation provides 19 model integrations across nine task families, covering language, speech, multimodal interaction, image/video generation, action policies, and world models (Appendix~\ref{appendix:model-coverage}). Deployment details appear in Appendix~\ref{appendix:deployment}.
\ifanonymoussubmission
Our implementation is publicly available.\footnote{\url{https://anonymous.4open.science/r/omniflow-anonymous-version-E75C}}
\else
Our implementation is publicly available.\footnote{\url{https://github.com/meituan-longcat/omni-flow}}
\fi

Across four Qwen3-Omni benchmarks with 128 samples each and concurrency levels
1, 2, and 4, Omni-Flow's aggregate JCT is 1.2\% higher than vLLM-Omni's and
4.9\% higher than SGLang-Omni's.  Its server-side conversational state removes
accumulated-history replay in multi-turn serving.  Over 50 turns, its mean JCT
is 7.5\% and 19.3\% lower than vLLM-Omni's and SGLang-Omni's, respectively;
over the last five turns, the reductions grow to 10.4\% and 28.6\%
(Section~\ref{sec:evaluation}).

\section{Related Work}
\label{sec:related-work}

\subsection{Inference Engines and Reusable State}

Inference systems optimize attention and execution~\cite{flashattention,deepspeed-inference}, KV paging and prefix reuse~\cite{vllm,sglang}, batching~\cite{orca,sarathi-serve}, and prefill/decode disaggregation~\cite{splitwise,distserve}. FlexGen~\cite{flexgen}, InfiniGen~\cite{infinigen}, and CacheGen~\cite{cachegen} address memory placement, KV prefetching, and compressed KV transfer, respectively. Omni-Flow reuses engine execution while separating storage ownership from scheduling, allowing compatible autoregressive and diffusion roles to access shared state under different execution policies.

CachedAttention~\cite{attentionstore}, Mooncake~\cite{mooncake}, and LMCache~\cite{lmcache} provide multi-turn cache management, distributed KV pools, and external cache layers. Preble~\cite{preble} schedules requests using prompt sharing and load; Llumnix~\cite{llumnix} migrates active request state across instances. Omni-Flow focuses on state access across heterogeneous roles: compatibility-aware KV reuse, intermediate tensors, and eligible checkpoint shards share a data plane whose storage lifetimes are independent of the consuming engines.

\subsection{Multimodal Serving Frameworks}

Recent multimodal serving frameworks extend LLM engines by decomposing model execution into stages. vLLM-Omni extends vLLM with stage-based execution for multimodal models~\cite{vllm-omni}. A pipeline configuration declares each model stage, its input sources, and bridge functions that translate outputs between engines. Requests can select where the configured pipeline terminates, and stage outputs can be streamed to the client or a downstream stage; this selection does not provide routing among arbitrary internal paths~\cite{vllm-omni-runtime}.

SGLang-Omni likewise decomposes a model into stages, while exposing \texttt{next}, \texttt{route\_fn}, \texttt{wait\_for}, and \texttt{stream\_to} in its pipeline configuration~\cite{sglang-omni-runtime}. These interfaces provide static transitions, request-dependent routing, fan-in, and stage-to-stage streaming. Its ordinary fan-in retains one payload slot per source stage.

Omni-Flow instead defines topology and field dependencies through a Python DSL. It targets workflows in which conditional paths can contribute repeated or request-dependent numbers of outputs, supporting request-dependent routing, cycles, causal grouping, runtime-declared counts, and execution-order constraints. Table~\ref{tab:capability-comparison} summarizes the native interfaces, and Appendix~\ref{appendix:workflow} gives representative definitions.

\begin{table*}[t]
\caption{Multimodal serving capabilities.}
\label{tab:capability-comparison}
\centering
\small
\renewcommand{\arraystretch}{1.08}
\setlength{\tabcolsep}{3pt}
\begin{tabular}{@{}lccc@{}}
\toprule
\textbf{Capability} & \textbf{vLLM-Omni} & \textbf{SGLang-Omni} & \textbf{Omni-Flow} \\
\midrule
\multicolumn{4}{@{}l}{\emph{Workflow coordination}} \\
Workflow definition & Pipeline + bridges & Config + routing functions & Python DSL \\
Fan-in / aggregation & Input-source tuple & One slot / source & Causal / count joins \\
Streaming output & \cmark & \cmark & \cmark \\
Request-dependent routing & \xmark & \cmark & \cmark \\
Cyclic workflow & \xmark & \xmark & \cmark \\
\midrule
\multicolumn{4}{@{}l}{\emph{State ownership and sharing}} \\
KV storage ownership & Stage / runner & Receiving stage & Shared data plane \\
Cross-process GPU weight sharing & \xmark & \cmark\ (replicas)$^{\dagger}$ & \cmark\ (role types)$^{\dagger}$ \\
\bottomrule
\end{tabular}
\par\vspace{1pt}\noindent
\begin{minipage}{0.84\textwidth}
\cmark: native support; \xmark: no native interface. $^{\dagger}$Same-device, compatible weights/layouts.
\end{minipage}
\end{table*}

The frameworks also differ in state ownership. vLLM-Omni supports cross-stage KV transfer and an experimental runner-owned session KV pool, whereas SGLang-Omni delegates destination-KV allocation and reclamation to the receiving stage~\cite{vllm-omni-runtime,sglang-omni-runtime}. Omni-Flow places configured physical KV pools under a shared data plane. For resident GPU weights, SGLang-Omni shares across co-located replicas of one logical process, whereas Omni-Flow shares compatible shards across role types on the same GPU. vLLM-Omni's host-weight reuse concerns CPU representations, with GPU transfer handled separately. Thus, Table~\ref{tab:capability-comparison} distinguishes ownership and GPU-sharing scope from data transfer alone.

\subsection{Dataflow and Workflow Runtimes}

Distributed termination detection~\cite{dijkstra-scholten} and Naiad~\cite{naiad} track outstanding work to establish progress. LangGraph~\cite{langgraph-runtime} provides bulk-synchronous orchestration; Ray~\cite{ray} and Dask~\cite{dask} provide distributed execution through actors and task graphs. Alpa~\cite{alpa} separates inter- and intra-operator parallelization for distributed training. Omni-Flow specializes coordination to inference with repeated yields, conditional delivery, and retained model state: inferred join scopes identify related contributions, leases establish completion, and declared predecessors constrain stateful execution. These mechanisms connect workflow coordination to shared-state access without imposing one model scheduler.

\section{Control Flow}
\label{sec:control-flow}

Multimodal applications compose model inference with application-specific preprocessing, input assembly, and output processing. Omni-Flow provides a declarative workflow programming model for composing these computations, including workflows with cycles. Users implement computational interfaces and connect their declared input and output fields through a Python DSL. The control-flow runtime manages dependency tracking and activation scheduling, separating interface computation from the coordination of asynchronous execution. Interfaces execute across workers and replicas, with related contributions routed to the same replica for aggregation; distributed execution and request-lifecycle details appear in Appendix~\ref{appendix:control-execution}.

Figure~\ref{fig:workflow-example} illustrates how these interfaces compose into a multimodal workflow. The parser emits text or image inputs: text follows an embedding path, while an image fans out to vision and VAE encoders whose results converge at generation. The LLM streams text back to the caller; its completion signal and the required conditioning inputs feed diffusion, followed by image decoding. This example brings together branch convergence, ordered updates to session state, and aggregation of request-dependent inputs. The following mechanisms define when these logical steps become ready and may execute.

These interfaces have different execution behaviors: some return a single result, others yield repeatedly, and still others aggregate a request-dependent number of inputs. A graph edge specifies where data flows, but does not by itself determine which outputs form one logical execution of an interface (an activation) or when that activation may execute. The numbered locations in Figure~\ref{fig:workflow-example} expose three synchronization problems.

\paragraph{Diamond convergence.}
At point~1, the vision and VAE encoder outputs converge after processing the same image along parallel paths. Their outputs must join at generation according to the parser occurrence from which they descend. More generally, a branch may yield repeatedly or spawn further activations, so receiving every required field does not establish completion. The join must identify both which contributions belong together and when no more can arrive.

\paragraph{Sequence dependencies.}
At point~2, successive parser frames traverse paths with different processing latencies, so the bucket for a later frame may become ready first. If generation updates the same session history, the later activation must wait for the preceding activation to finish, preserving the parser's yield order in history updates. Likewise, tokens from one generation may arrive at the entry node out of order; the entry node uses their explicit predecessor relationships to restore the generation order before presenting them to the user.

\paragraph{Dynamic-count joins.}
At point~3, the diffusion stage waits for the LLM completion signal and a request-dependent number of vision embeddings and VAE latents. It must collect the declared count for each field, together with associated metadata. Unlike the per-image diamond join, this stage collects conditioning inputs from multiple parser outputs within one request, using runtime-supplied counts to determine when all required inputs have arrived.

Omni-Flow addresses these patterns by separating \emph{causal grouping}, \emph{contribution closure}, and \emph{execution ordering}. Token paths and statically inferred scopes associate diamond contributions (Section~\ref{sec:three-deps}); causal leases establish their closure (Section~\ref{sec:contribution-closure}). Explicit predecessors constrain contribution admission and stateful execution (Section~\ref{sec:control-order}). Count joins instead form request-wide input sets using declared counts and associated fields (Section~\ref{sec:count-join}). Within the supported graph structures, these mechanisms handle combinations of the three patterns through a common node-local scheduling protocol.

\begin{figure}[t]
    \centering
    \includegraphics[width=\columnwidth]{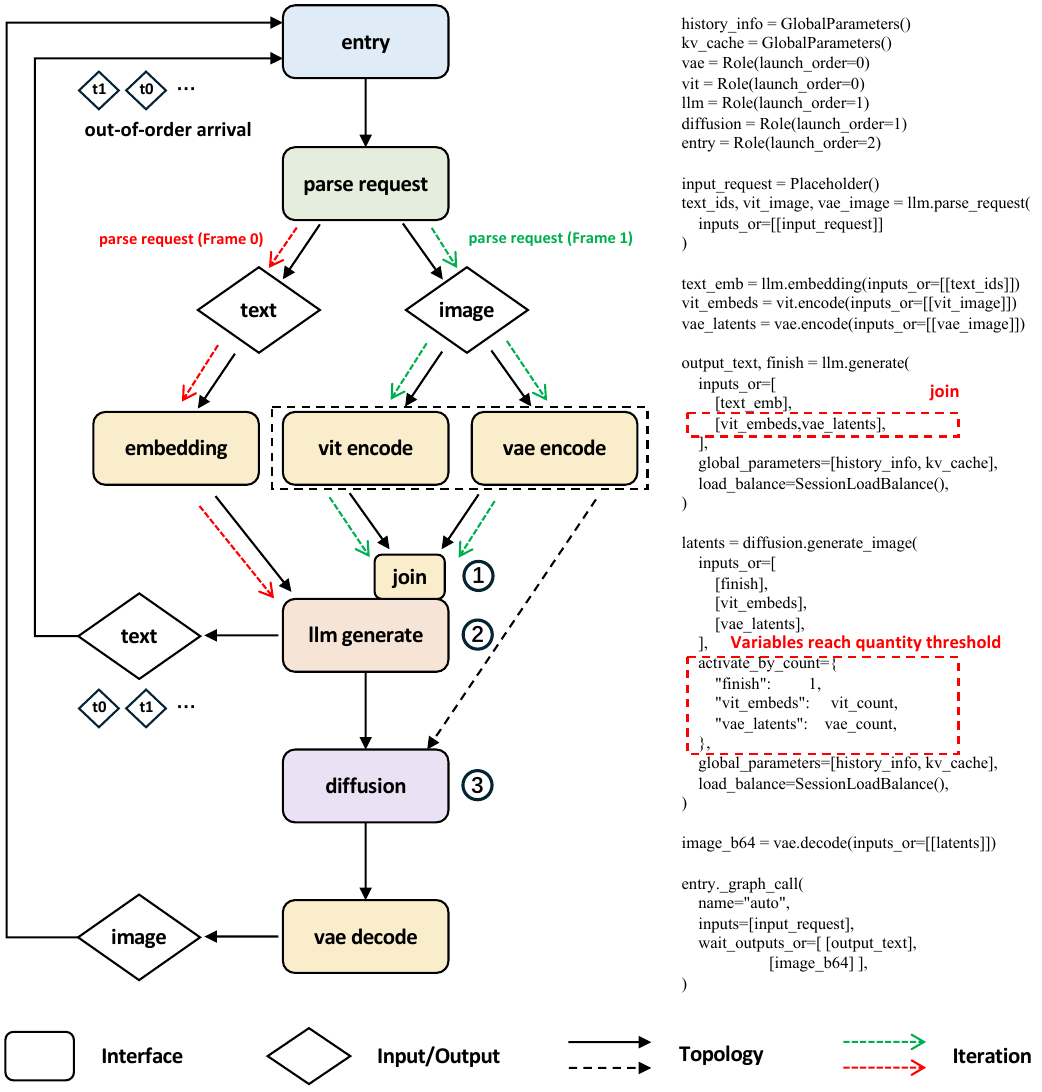}
    \caption{Workflow and DSL example.}
    \label{fig:workflow-example}
\end{figure}

\subsection{Interface Definitions and Graph Planning}
\label{sec:graph-parsing}

In the DSL, each computational interface forms a logical graph node, and field mappings between interfaces form directed edges. An interface declares the fields required for one execution and may provide several independent input alternatives. Placeholders supply graph-call inputs, while global parameters connect interfaces to session-scoped state managed by the data plane. These declarations describe logical dependencies independently of the physical instances that execute them.

Before serving requests, Omni-Flow's control-flow runtime parses the workflow definition to determine where each interface's inputs come from, which fields each route can provide, and when the interface has received enough input to execute. An interface may be triggered by one complete upstream output, may need to combine fields arriving along several paths, or may collect a request-dependent number of values. If an output cannot provide any field required by a route, the runtime marks that delivery as skipped, allowing the receiver to distinguish an absent input from one that is still in flight.

When inputs arrive through several paths, graph parsing also identifies their common upstream origin. The runtime uses this information to combine results belonging to the same logical execution even when they arrive out of order; it never infers their relationship from arrival order. If an input dependency cannot be resolved unambiguously, the control-flow runtime reports a configuration error during initialization. Cyclic workflows are supported as long as contributions from different iterations remain distinguishable. The following sections describe how the runtime groups these inputs, determines that no more can arrive, and enforces their required execution order.

\subsection{Tracking and Grouping Execution Origins}
\label{sec:three-deps}

\begin{figure}[t]
  \centering
  \includegraphics[width=\columnwidth]{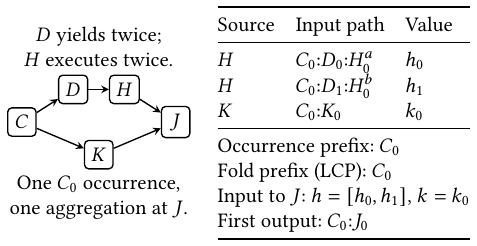}
  \caption{Causal grouping and join completion. Superscripts identify activations; subscripts identify yields.}
  \label{fig:control-causality}
\end{figure}

Figure~\ref{fig:control-causality} shows the grouping problem. Under source output $C_0$, $D$ yields twice and consequently invokes $H$ twice. The resulting values $h_0$ and $h_1$, together with $k_0$ from the other branch, all belong to the same logical execution at $J$. Results descended from the next source output $C_1$ belong to a different execution and must not be mixed with them. The control-flow runtime must preserve this grouping even when results arrive out of order. This section explains how Omni-Flow identifies which results belong together; Section~\ref{sec:contribution-closure} explains how it knows that all such results have arrived.

Each yielded output carries a token path recording its execution origin. Graph planning identifies the common upstream origin, or \emph{anchor}, of a multi-path join. The path through a particular anchor output identifies one occurrence: results descended from $C_0$ enter one bucket, while those from $C_1$ enter another. Buckets are scoped to the graph request, receiving interface, and input group.

After the join executes, the runtime retains the longest common prefix (LCP) of its input paths and appends the new output token. In Figure~\ref{fig:control-causality}, this produces $C_0{:}J_0$, preserving the common origin for downstream grouping. Bucket identity remains fixed while contributions arrive, even if their current LCP changes. Appendix~\ref{appendix:prefix-join} specifies the identifiers and path operations.

\subsection{Contribution Closure Detection}
\label{sec:contribution-closure}

Section~\ref{sec:three-deps} determines which inputs belong to a bucket, but having all required fields does not mean that no more inputs will arrive. In Figure~\ref{fig:control-causality}, $h_0$ and $k_0$ already cover the fields required by $J$, yet $D$ may yield again and cause the second execution of $H$ to produce $h_1$. Invoking $J$ at this point would omit a later result that belongs to the same bucket.

To prevent such premature execution, the control-flow runtime tracks all upstream work that may still produce inputs for each bucket. Before a task dispatches child work that can produce another input, it registers that relationship. Tracking for the task finishes only after the task can create no more children and every registered child has also finished. Omni-Flow implements this recursively expanding tracking relationship with \emph{causal leases}; registering children before dispatch prevents a bucket from appearing complete while new work is being created.

In Figure~\ref{fig:control-causality}, $C_0$ creates an independent bucket, so that bucket does not wait for the later source output $C_1$. Within the $C_0$ bucket, however, the runtime continues waiting while $D$ may yield again, and both executions of $H$ are included in the tracked work. The bucket becomes ready only after all work that can still contribute to it has finished and its collected fields satisfy $J$'s input requirements. If the tracked work finishes but required fields are still missing, the control-flow runtime marks the bucket as impossible instead of continuing to wait or invoking $J$. Appendix~\ref{appendix:prefix-join} further describes grouping, contribution closure, and input admission when source and intermediate nodes yield repeatedly.

\subsection{Contribution Order and Stateful Execution}
\label{sec:control-order}

Asynchronous transport may cause consecutively produced results to arrive out of order. Omni-Flow therefore preserves two kinds of order: the order in which results enter an ordinary multi-path bucket, and the execution order of interfaces that update shared session state. The former preserves upstream yield order when inputs are merged; the latter preserves input order when session state is updated.

For an ordinary multi-path bucket, a contribution waits until its explicit predecessor has committed. In Figure~\ref{fig:control-causality}, $h_1$ therefore waits for $h_0$ even if it arrives first, so $J$ receives $h=[h_0,h_1]$. The parallel $K$ branch has no additional ordering relation with $H$. Token paths determine bucket membership; explicit predecessors determine contribution order. Appendix~\ref{appendix:prefix-join} gives the admission rule and treatment of skipped deliveries.

Interfaces declared with \texttt{SessionLoadBalance} that update shared session state also require execution ordering. For example, a parser may produce $C_0$ and then $C_1$. If the paths derived from $C_1$ complete sooner, the corresponding bucket may satisfy the activation conditions of the generation interface first. If both generations update the same session history, however, the execution for $C_1$ must wait until the execution for $C_0$ has settled.

\subsection{Count Joins for Dynamic Input Sets}
\label{sec:count-join}

Some interfaces need a request-wide input set whose size is known only at runtime. Point~3 in Figure~\ref{fig:workflow-example} shows this case. Before invoking diffusion, the workflow must collect one LLM completion signal, \texttt{vit\_count} vision embeddings, and \texttt{vae\_count} VAE latents. These values may come from different parser outputs and arrive in any order. Diffusion can execute only after the declared count of every required field has arrived.

An \texttt{activate\_by\_count} (ABC) join aggregates inputs within a request according to declared counts; these inputs may come from different parser outputs. Fields listed in \texttt{activate\_by\_count} are trigger fields. The interface can execute once each trigger field has the declared number of distinct events and its associated fields are available; it fires at most once per graph call.

Readiness depends on these counts and associated fields, without waiting for prefix-lease closure or using the per-occurrence execution gate. Selection precedes path sorting, and the output retains the selected paths' structural LCP. Appendix~\ref{appendix:count-contract} details event selection, companion matching, zero counts, and ordering.

\section{Data Flow}
\label{sec:data-flow}

Control flow (Section~\ref{sec:control-flow}) schedules multimodal roles independently; data flow connects these roles through shared model weights, reusable KV state, and intermediate-tensor exchange. In unified understanding--generation models such as BAGEL and HunyuanImage~3.0~\cite{bagel,hunyuanimage3}, matching checkpoint parameters can be shared across role instances to reduce GPU memory use, while existing context KV can be reused during generation to avoid repeated context computation.

\begin{figure*}[t]
    \centering
    \includegraphics[width=\textwidth]{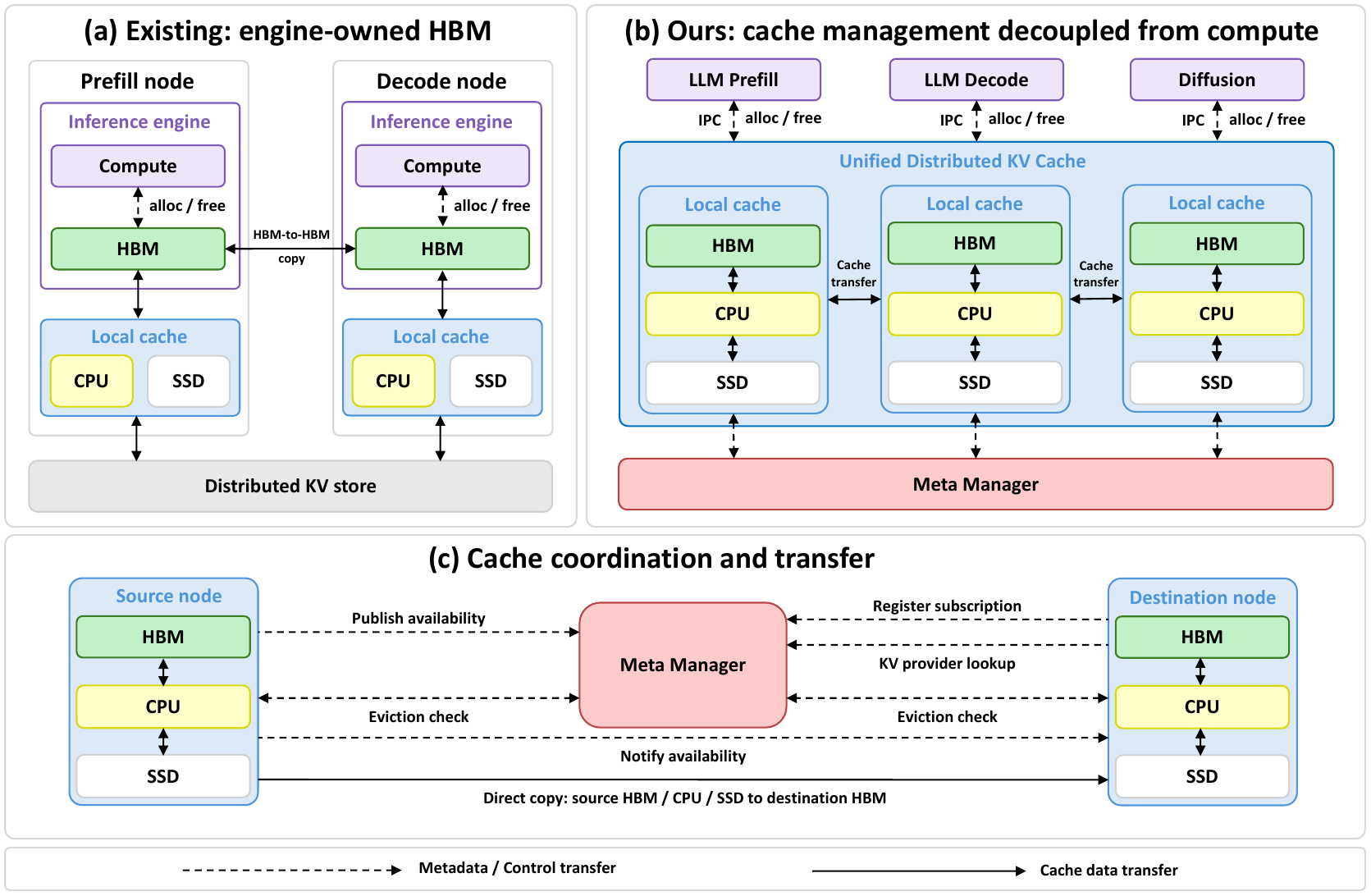}
    \caption{Cache organization in existing engines and Omni-Flow.}
    \label{fig:data-flow}
\end{figure*}

In conventional serving systems, KV ownership follows the inference-engine boundary (Figure~\ref{fig:data-flow}(a)). Each engine allocates and reclaims its own GPU cache, while prefill/decode transfer, local CPU or SSD offloading, and cross-service cache reuse are commonly handled by separate mechanisms. These mechanisms work well for their respective settings, but composing several independently scheduled roles requires additional pairwise transfer paths and lifecycle coordination.

Omni-Flow instead separates cache management from model execution through a shared data plane (Figure~\ref{fig:data-flow}(b)). The data plane manages physical allocation, tier residency, provider discovery, transfer, and reclamation, while compute adapters retain model-specific prefix and update semantics. KV lookup and transfer are determined by page identity and layout compatibility, so the system does not require a separately configured cache path for each pair of upstream and downstream roles. The same interface therefore connects KV produced and consumed by different compatible roles or replicas, while local weight sharing and intermediate-tensor exchange use storage suited to their own lifetimes.

The data plane follows a client--server architecture. A node-local \texttt{MemoryManager} manages registered pools, while role processes access their assigned manager through \texttt{MemoryManagerClient} over ZMQ~\cite{zeromq} IPC. The subsystem provides three complementary mechanisms for different lifetimes and sharing requirements:

\begin{itemize}[itemsep=3pt,topsep=2pt,parsep=0pt,partopsep=0pt]
    \item \textbf{Global Params Pool:} The designated destination for the KV cache, using a paged, tiered storage hierarchy (L1 GPU / L2 CPU / L3 SSD) with automatic data migration between tiers based on hot and cold access patterns.
    \item \textbf{Tmp Buffer Pool:} The transit hub for intermediate data, embedded in the system to automatically provide zero-copy, low-latency transfer of input and output tensors across role interfaces.
    \item \textbf{Weight Sharing:} An IPC-backed registry through which co-located roles reuse compatible checkpoint parameters on the same device, while retaining role-specific parameters and runtime state separately.
\end{itemize}

\subsection{Global Params Pool}

Unlike existing KV-cache schemes, in which inference engines manage their own cache structures, dedicated channels transfer KV between fixed stages, and local lower-tier storage is managed separately, the Global Params Pool brings these responsibilities under one data plane. Its core differences are:

\begin{itemize}[itemsep=3pt,topsep=2pt,parsep=0pt,partopsep=0pt]
    \item \textbf{Unified allocation and reclamation.} The data plane allocates and reclaims physical slots in registered KV pools. Inference engines remain responsible for model execution, while compute adapters preserve model-specific prefix and update semantics and translate data-plane slots into engine indices.
    \item \textbf{Unified tier and lifecycle management.} Node-local managers coordinate residency across the configured GPU, CPU, and SSD tiers, while distributed metadata records layouts, session page chains, and providers. Publication and subscription allow consumers to preload reusable pages before access.
    \item \textbf{Layout-based provider discovery.} For a requested page, the data plane finds available providers whose layouts jointly cover the required representation. Provider selection is based on page identity and compatible layout, rather than a fixed role or replica pairing.
\end{itemize}

The subsystem is organized as a physical paging layer, a multi-tier slot engine, distributed metadata coordination, and a complete data-access lifecycle.

\subsubsection{Physical Paging Layer}

The Physical Paging Layer provides low-level storage management for the configured GPU, CPU, and SSD tiers and exposes the tensor regions used by the Slot Engine. Rather than imposing one monolithic tensor layout, each global parameter describes named outer dimensions (\texttt{sub\_names}, such as Transformer layers), named slice dimensions (such as K/V or head partitions), and its page size, shape, and dtype. Upper layers retrieve tensor views or region descriptors using $(\mathit{sub\_names\_tuple},\,\mathit{page\_idx},\,\mathit{slice\_idx\_tuple})$. This compositional description lets heterogeneous models reuse the same paging interface without changing allocation code for each tensor layout. The returned views expose existing storage without copying; moving or assembling data across storage locations remains a separate operation.

\subsubsection{Multi-Tier Slot Engine}
\label{sec:slot-engine}

The Slot Engine maps logical pages to physical slots across the configured tiers. Eviction is guided by a TinyLFU-inspired frequency sketch~\cite{tinylfu} with periodic decay and a four-bucket LRU (cold / warm1 / warm2 / hot), which selects colder candidates first. Because a page may simultaneously be used by inference or by loading, offloading, and uploading operations, each slot tracks these four kinds of holders in separate identifier sets. Repeated acquisition by the same identifier is idempotent, and a slot can be reclaimed locally only when all four sets are empty.

Cross-tier migration converts between compute-oriented layouts, including NHD block stacks, and the representation used by lower-tier storage. Supported layouts use CUDA kernels to copy and convert pages directly into destination buffers without an additional staging allocation; other layouts use a PyTorch fallback. Concurrent requests for the same page join the same in-flight upload and reuse its result, avoiding duplicate transfers.

\subsubsection{Distributed Metadata Coordination}

The Distributed Metadata Coordination Layer uses Redis~\cite{redis} to coordinate shared layout, session, and provider metadata. Lua scripts make critical metadata transitions atomic, while pipelines batch queries and updates. Redis identifies where compatible page representations are available; physical slots and active I/O remain under node-local management. The layer is organized around four functions.

\textbf{Layout Identification and Coverage.} Each layout receives a deterministic identifier derived from canonical layout metadata. A provider may expose one or more layout shards. When a complete page spans several layouts, the system enumerates candidate groups whose union covers all required components and prefers groups with fewer layouts. The layout identifier describes a page's representation independently of its content identity.

\textbf{Session Prefix Matching.} Given a session and an ordered page query, the metadata layer finds the reusable prefix in the stored page chain. When provider-coverage validation is enabled, matching stops at the first page for which no complete candidate group is available. Callers may also truncate the session's visible page range, while model-specific construction of the query remains in the compute adapter.

\textbf{Replica Publication and Provider Selection.} After computation, publication registers a node as a provider of the completed page layouts. A query selects an available candidate group for each page, preferring one provider that covers the whole group when possible and otherwise gathering the required layouts from multiple providers.

\textbf{Protected Metadata Eviction.} Before removing provider metadata, one Lua operation checks provider-use locks, session-running state, and the applicable layout-coverage requirement. When preservation is required, at least one complete candidate group must remain; whether the last provider may be removed depends on the parameter's recomputation policy. This atomic operation protects metadata removal, while copy completion and physical slot reclamation remain protected by the local Slot Engine.

\subsubsection{Complete Data Flow Loop}

The complete data-access loop connects these layers as data is produced, exposed to consumers, acquired, and eventually retained or reclaimed. Global parameters and temporary buffers follow different retention policies after all active references are released. For global KV pages, Figure~\ref{fig:data-flow}(c) separates metadata operations---subscription, provider lookup, and eviction checks---from direct availability notifications and KV transfer between node-local memory managers.

\textbf{Initialization.} When the upper layer creates a global parameter, the API allocates or reuses its physical paging storage and registers newly allocated memory regions with the transfer engine. Pool finalization establishes the configured L1/L2/L3 storage hierarchy and registers the node's layout group and candidate combinations in Redis.

\textbf{Inference Write $\rightarrow$ Publication $\rightarrow$ Preload.} Upon completing computation, a request invokes the publication interface, which maps page IDs to local slots and registers the node as a replica provider. When session retention is enabled, it also updates the session's page chain. After the metadata update is acknowledged, the publishing memory manager delivers batched availability notifications to registered subscribers to trigger downstream preload.

\textbf{Cross-Node Demand Transfer via RDMA.} On a preload notification or local cache miss, the consumer first searches its local multi-tier storage, then queries Redis for providers of missing components. Transfer follows a Search--Transfer--Release lifecycle: the selected provider pins the required source pages, the consumer supplies destination-buffer descriptors, and the provider issues batched RDMA writes. After a successful transfer, source-page pins are released; uncertain completion leaves the affected buffers protected.

\textbf{Copy-on-Write and Partial Page Handling.} When incremental data extends a retained partial page, the system copies its existing contents into a writable slot and appends the new data there. The updated page is published under a new page ID, preserving the old published identity even when only one request holds it. For default session page IDs, dirty keys distinguish partial writes, while session metadata records the total valid token length. Retention follows the parameter's recomputation policy: for non-recomputable state, eviction must preserve an available complete representation, since a consuming role may be unable to regenerate another role's outputs.

\textbf{Eviction Under Memory Pressure.} When local HBM is under pressure, the Slot Engine selects cold-priority candidates and coordinates migration or removal with protected metadata checks in Redis. These checks cover provider-use locks, session-running state, and the required layout coverage under the parameter's recomputation policy. The four reference sets in Section~\ref{sec:slot-engine} prevent reclamation while requests or I/O operations hold a slot. Any required transfer and metadata update must complete before the slot is reused; requests may wait or fail if no safe capacity is available.

\subsection{Tmp Buffer Pool}

Unlike KV pages retained for later reuse, intermediate tensors such as activations and encoder outputs usually pass between a few interfaces within one request. An ordinary temporary block is reclaimed when its references reach zero; only a published cache block remains available for reuse. The Tmp Buffer Pool gives these short-lived values request-scoped ownership: consumers on the same device can share storage, while consumers on another device or node receive a copy in their selected location. Intermediate tensors are allocated from preallocated, byte-addressable pools. Each pool suballocates variable-sized blocks from its backing storage, avoiding a fresh backing allocation for every intermediate tensor.

\textbf{Reference Counting and Cached Reuse.} Each new block starts with one reference; a same-device consumer can acquire another reference without copying the tensor. Releasing the last reference reclaims an ordinary block, whereas a block still registered under a cache key is retained for reuse. A producer explicitly publishes a completed block under a cache key. Subsequent requests borrow it by incrementing its reference count, and the LRU policy evicts only idle entries whose reference count is zero. An offset index resolves interior tensor pointers to their owning allocations so that sliced tensor descriptors retain whole-block reference accounting.

For cross-host intermediates, the receiver materializes tensors from received descriptors by allocating local storage and issuing batched RDMA reads from the producer's registered buffers. The consumer selects the destination device. Same-device reuse avoids a copy when the allocation is shareable; other local device placements require a copy.

\subsection{Weight and KV Cache Sharing}

\begin{figure}[t]
    \centering
    \includegraphics[width=\linewidth]{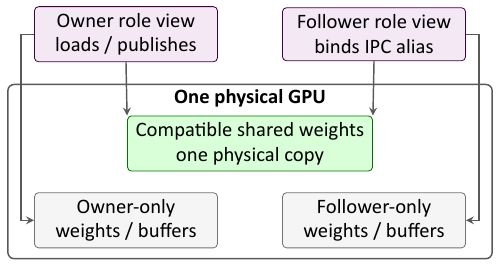}
    \caption{Same-device weight sharing across roles.}
    \label{fig:weight-sharing}
\end{figure}

\textbf{Model Weight Sharing.} When independently scheduled roles run as separate processes, their compatible parameters would normally be allocated and loaded repeatedly. Co-located roles under the same memory manager instead share compatible checkpoint parameters through an IPC-backed registry. Shared entries are named by checkpoint identity, parameter name, and TP/EP partition metadata. Initialization is serialized per device: the owner constructs and loads the model, performs post-processing, and publishes identified checkpoint parameters. Followers construct parameters on the meta device and bind matching entries to shared tensors. In the standard loading path, unmatched parameters receive local storage and may be published after loading. A \texttt{SkipWriteParameter} wrapper suppresses supported repeated in-place updates during follower loading. Clients poll loading entries for readiness instead of occupying server workers while waiting. After initialization, shared weights occupy one physical copy per device within the manager; role-specific parameters and runtime buffers retain separate storage (Figure~\ref{fig:weight-sharing}).

\textbf{KV Cache Sharing.} A co-located role that measures free memory before its peers finish loading weights can overestimate the capacity available to a shared KV pool. Participants in the same named pool under a memory manager therefore negotiate a common logical page capacity before physical allocation. A launch-planned device barrier first lets participating roles finish loading weights before measuring free device memory. The page count is then finalized as the minimum of the participants' proposals, so that the pool's physical shards or replicas agree on a supported capacity. Compatible LLM and diffusion roles can use this mechanism to share KV storage. Shared storage alone does not make KV values from different models semantically interchangeable.

\section{Compute Flow}

Control Flow and Data Flow provide workflow orchestration and shared-state
management; Compute Flow connects model-specific execution to these
abstractions.  Multimodal workloads combine different encoders, backbones,
decoders, and iteration patterns.  Omni-Flow accommodates them through
model-specific adapters over common execution and memory interfaces.  The
adapters integrate replayable multimodal inputs, framework-managed KV, and
model-specific sampling hooks with SGLang~\cite{sglang}; compatible diffusion
backbones reuse its transformer execution interfaces under separate
scheduling.

\begin{figure*}[t]
  \centering
  \includegraphics[width=\textwidth]{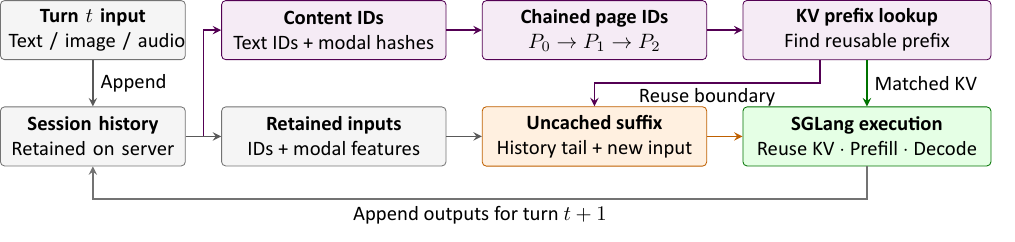}
  \caption{Multimodal prefix matching in Omni-Flow.}
  \label{fig:prefix-history-comparison}
\end{figure*}

\subsection{Multimodal Prefix Matching}

Traditional LLM prefix caching matches token-ID sequences.  This is
insufficient for multimodal histories: the same modality placeholder can refer
to different images or audio, while embeddings and model-specific auxiliary
fields needed for recomputation cannot be recovered from that placeholder.
Omni-Flow therefore separates the identity used to match a prefix from the
model inputs used to execute it.

\textbf{Complete-Input Preparation.}
After the relevant encoders finish, the join path assembles aligned token IDs
and content hash IDs, together with modality embeddings and auxiliary fields
when required.  Text positions retain their token IDs, whereas modal positions
carry content-derived identities and are marked as externally embedded.
Eligible encoder outputs may be reused through a cross-request LRU cache whose
key covers the payload, preprocessing options, and model context.  Input
assembly does not itself claim a KV hit: the generation adapter performs that
lookup at execution time and restricts it to the pre-existing-history boundary.

\textbf{Incremental Session Updates.}
A request can append only the current turn to server-retained multimodal
history.  An explicit nonnegative \texttt{step} first caps the retained logical
history at that token position; an absent or negative value preserves the
current history before appending.  The merge path stores the new input run and
reports both the resulting prefill length and the boundary preceding the new
run.  Generated tokens and configured history fields are retained so that the
next turn can continue from the updated session.

Figure~\ref{fig:prefix-history-comparison} summarizes this path.  Content
identities resolve reusable KV, while retained inputs supply the unmatched
suffix to SGLang for prefill and decoding.  Retained outputs then extend the
next turn's history.

Both paths use \textbf{Chained Paged Hashing}: each page ID hashes its
predecessor together with the current page's content identities, thereby
representing the complete page-aligned prefix.  The generation adapter queries
global KV providers with these IDs independently of the session-history page
IDs.  Matching stops at the first absent or incompatible page; cross-session
or cross-replica reuse also requires consistent page conventions and
compatible KV representations.

Prefix reuse is decomposed into two semantic layers.
\textbf{Replayable Inputs} are retained, when configured, in the CPU-backed
\texttt{history\_info} pool: token IDs, content hash IDs, modality embeddings,
and auxiliary fields required by the model.  They allow an unmatched suffix to
be reconstructed and prefilled when the required fields remain available.
\textbf{Materialized KV} is resolved separately at execution time and reused
only when a compatible provider is available.  This separation makes KV reuse
opportunistic without making correctness depend on KV residency.

Input preparation and KV resolution occur at different boundaries.  Merge and
join paths update or assemble logical history; the generation adapter then
determines the model-valid reuse boundary, maps matched pages to physical
slots, and includes the remaining suffix in prefill.  Newly joined content is
therefore never treated as already cached.

\subsection{Interface Abstraction over SGLang and KV Takeover}

A process-local request-cache adapter connects model I/O to SGLang.  Request
cache I/O and external KV ownership are controlled independently.  With
external KV enabled, Omni-Flow owns the registered storage, while SGLang
retains scheduling, request-to-token mappings, and attention execution.
Model-specific hooks can return auxiliary tensors while invoking SGLang's
native sampler.

\textbf{Abstracted I/O Parameter Interface.}
Requests carry session and sequence metadata, generation limits, and physical
slots on the external-KV path.  Configured I/O keys, tensor schemas, and
model-declared specifications define the forward-batch interface.  Inputs may
arrive directly or from retained history.  Ordinary forwards use batch-owned
buffers; CUDA Graph replay uses fixed-address staging buffers.  Produced
tensors, sampled tokens, and configured fields are retained when their output
or history storage is bound.

\textbf{External KV-Cache Ownership.}
When enabled, the SGLang cache adapter bypasses native radix-tree insertion and
maps framework-supplied slots and prefix lengths to engine KV indices.  The
data plane allocates, retains, and publishes KV pages; SGLang preserves engine
index conventions and request mappings, and its attention kernels use the
assigned slots.  Before releasing a request-pool row, the completion hook waits
for queued page publication.  Retention and reclamation then follow the
data-plane lifecycle.

\subsection{Diffusion Reusing Transformer Execution}

Diffusion Transformers perform iterative denoising with transformer
backbones~\cite{dit,ddpm,latent-diffusion}.  Compatible backbones reuse
SGLang's transformer execution and memory interfaces, but not its
autoregressive scheduling loop.  The diffusion runtime schedules requests and
invokes a resident SGLang \texttt{ModelRunner} through external-forward,
supplying sequence lengths, physical KV slots, latents, timesteps, and
attention metadata.  This preserves diffusion-specific scheduling and
attention semantics while reusing paged-attention and weight infrastructure.

The motivation is parallelism alignment.  A common transformer backend lets
compatible LLM and diffusion roles use aligned tensor- and expert-parallel
parameter layouts.  Physical weight sharing still requires matching parameter
identities and shard layouts; compatible shards on each GPU can then reference
storage managed by the same MemoryManager.  KV reuse likewise requires
compatible cache semantics, layouts, and content-page identities.  For
example, BAGEL's generation expert can attend to context KV produced by its
understanding expert despite using different projection parameters~\cite{bagel}.
When these conditions hold, a diffusion role can reuse immutable conditioning
KV while keeping changing denoising state in private writable pages.

\section{Evaluation}
\label{sec:evaluation}
\definecolor{evalblue}{RGB}{76,105,230}
\definecolor{evalgray}{RGB}{145,145,145}
\definecolor{evallightgray}{RGB}{182,182,182}

By comparing Omni-Flow with vLLM-Omni and SGLang-Omni, we evaluate whether it
(1) preserves comparable end-to-end performance while executing Qwen3-Omni
through its general workflow abstractions and (2) benefits stateful, growing
conversations. We also evaluate control-flow scheduling, data-flow resource
management, and compute-flow history-metadata management overheads
(Appendix~\ref{appendix:overhead}).

\subsection{Experimental Setup}

\paragraph{Model and workload.}
We evaluate Qwen3-Omni-30B-A3B-Instruct on the first 128 examples from four
benchmarks that exercise different input modalities: RealWorldQA (image and
text)~\cite{realworldqa}, VoiceBench/OpenBookQA (audio and
text)~\cite{voicebench}, the \emph{speech} subset of
OmniBench (audio, image, and text)~\cite{omnibench}, and Daily-Omni (video,
separate audio, and text)~\cite{daily-omni}.  All three systems process the
same examples in source order, use greedy text decoding with at most 10
generated tokens and seed 42, and generate 16 valid codec frames.  Daily-Omni
video is sampled at 2 fps with at most 20 frames.

\paragraph{Systems and placement.}
All three systems use identical BF16 weights without quantization on a server
with two NVIDIA H20-3e GPUs, an Intel Xeon Platinum 8480+ CPU, and 392 GiB of
memory.  GPU~0 hosts the Thinker and multimodal encoders; GPU~1 hosts the
Talker and audio decoder.

\paragraph{Metrics and protocol.}
All timing is measured at the streaming client.  \emph{Job completion time}
(JCT) spans request submission through the terminal response.  For each
benchmark and concurrency, we restart each system, warm it with three full
concurrency waves, and report the mean of three runs over the same 128 requests.
All systems generate 16 codec frames; RTF is therefore proportional to JCT,
so we report only JCT.

\begin{figure}[t]
  \centering
  \makebox[\columnwidth][c]{\hspace*{-6pt}%
    \includegraphics[width=\columnwidth]{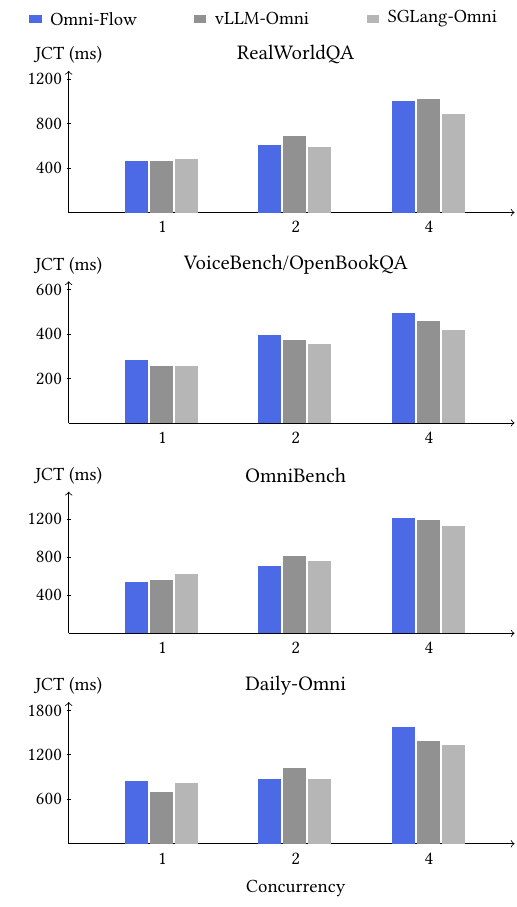}\hspace*{6pt}}
  \caption{JCT across benchmarks.}
  \label{fig:online-e2e}
\end{figure}

\subsection{Online End-to-End Performance}

Figure~\ref{fig:online-e2e} reports mean client-observed JCT.  We summarize a
set $\mathcal{C}$ of benchmark--concurrency configurations using:
\begin{equation}
  \Delta_s(\mathcal{C}) =
  \left(\prod_{(b,c) \in \mathcal{C}}
  \frac{J_{\textsc{Omni-Flow}}(b,c)}{J_s(b,c)}\right)^{1/|\mathcal{C}|}-1,
  \label{eq:aggregate-jct-gap}
\end{equation}
where the exponent is $1/|\mathcal{C}|$.  For each benchmark, $\mathcal{C}$
contains its three concurrency configurations.  Omni-Flow's JCT is 4.4\% lower
than vLLM-Omni's and 3.9\% higher than SGLang-Omni's on RealWorldQA; 8.7\% and
13.9\% higher on VoiceBench/OpenBookQA; 4.6\% and 4.2\% lower on OmniBench;
and 5.9\% and 6.9\% higher on Daily-Omni, respectively.  Thus, Omni-Flow leads
both baselines on OmniBench, trails both on VoiceBench/OpenBookQA and
Daily-Omni, and falls between them on RealWorldQA.  Across all 12
configurations, Omni-Flow's aggregate JCT is 1.2\% higher than vLLM-Omni's and
4.9\% higher than SGLang-Omni's.  Omni-Flow's aggregate single-turn JCT is
slightly higher, part of which comes from maintaining server-side session
state.  We consider this overhead necessary for stateful
multi-turn serving, where its benefits grow with conversation length
(Section~\ref{sec:eval-multiturn}).

\subsection{Stateful Multi-Turn Serving}
\label{sec:eval-multiturn}

We evaluate a 50-turn session using the first 50 OmniBench speech samples.
Each turn adds a distinct image, audio clip, and question.  Images are resized
to at most 896 pixels on their long edge; audio is mono 16-kHz and at most five
seconds long.  Generation is capped at 10 text tokens and fixed at 16 valid
codec frames (1.257 seconds of audio).  All systems reuse cached Encoder,
Thinker, and Talker history, while Code2Wav is not cached across turns.

\begin{figure}[t]
  \centering
  \includegraphics[width=\columnwidth]{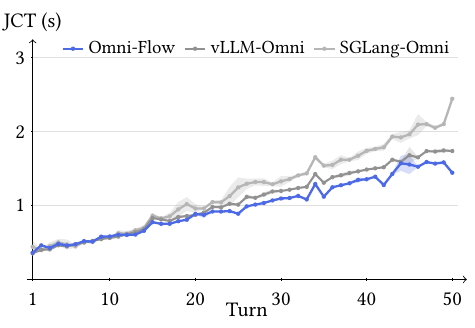}
  \caption{Fifty-turn JCT (mean $\pm$ SD).}
  \label{fig:multiturn-jct}
\end{figure}

Omni-Flow sends only the new turn and reuses server-side session state.  The
baselines replay the accumulated messages from the client.  We restart each
system for three runs, warm each run with the 51st sample, and then execute all 50 turns.
Figure~\ref{fig:multiturn-jct} reports the per-turn mean JCT; bands show
$\mathbin{\pm}1$ sample standard deviation across runs.

Across the three runs, mean JCT over all 50 turns is 980.6 ms for Omni-Flow,
compared with 1060.2 ms for vLLM-Omni and 1214.4 ms for SGLang-Omni, reducing
mean JCT by 7.5\% and
19.3\%, respectively.  Over the last five turns, the reductions grow to 10.4\%
and 28.6\%.  All systems slow as history grows: prefix caching avoids
recomputing old KV and Encoder outputs, but new tokens still attend to a longer
KV history, while client-side replay additionally reconstructs accumulated
messages and multimodal references.  Stage timings show that Code2Wav is not
the main source of the widening gap.

\section{Conclusion and Future Work}

Omni-Flow separates workflow orchestration, shared-state management, and model execution for heterogeneous multimodal inference. Control flow defines heterogeneous workflows through a Python DSL, while the runtime schedules components from their dependencies. Data flow manages the placement, transfer, and lifetime of shared tensors and paged KV caches, with same-device weight sharing. Compute flow connects model execution to the shared data plane through multimodal history matching and SGLang adapters, enabling KV reuse across turns and compatible roles. Omni-Flow supports 19 model integrations across nine task families. On Qwen3-Omni, its aggregate JCT across four benchmarks and concurrency levels 1, 2, and 4 is 1.2\% and 4.9\% higher than vLLM-Omni's and SGLang-Omni's, respectively; in a 50-turn multimodal session, it reduces mean JCT by 7.5\% and 19.3\%.

While continuing to optimize performance, future work will investigate workload-aware routing and placement that account for reusable KV, transfer cost, and queueing delay, together with resource coordination among co-located roles under changing load. We will also extend KV interoperability to additional attention and parallelism layouts and study latency, throughput, and memory tradeoffs under sustained multimodal interaction.

\bibliographystyle{ACM-Reference-Format}
\bibliography{omniflow}
\clearpage
\appendix

\begin{center}
{\large\textbf{Appendix}}
\end{center}
\vspace{0.5em}

\section{Additional Control-Flow Semantics}
\label{appendix:nested}

\subsection{Prefix-Join Semantics}
\label{appendix:prefix-join}

\paragraph{Bucket identity and output paths.}
To retain execution origin, each yielded output carries the path through which it was produced. At every yield, the runtime appends a token identifying the node, activation, yield position, and event to the inherited path. An activation may yield several outputs, and one output may be delivered to several downstream interfaces; delivery identities also allow receivers to suppress retransmissions.

When results from several paths meet, graph parsing has already identified their common upstream origin, called the \emph{anchor}. The path through the corresponding anchor output forms an \emph{occurrence prefix}, $P_o$. The receiver uses this prefix to place an arriving result into a stable bucket:
\[
  B = (r, i, g, \operatorname{hash}(s,P_o)),
\]
Here, $B$ is the bucket identifier. $r$ identifies one graph request using its session and graph-call identifiers, so separate graph requests in the same session cannot be mixed. $i$ is the receiving interface ($J$ in Figure~\ref{fig:control-causality}); $g$ is one input group declared through that interface's \texttt{inputs\_or}; $s$ identifies the multi-path merge derived from the graph; and $P_o$ identifies the particular upstream occurrence, such as $C_0$. The values $h_0$, $h_1$, and $k_0$ have the same values for all five components and therefore enter the same bucket. Results descended from $C_1$ have a different $P_o$ and enter a different bucket.

Once the runtime determines that a bucket is ready (Section~\ref{sec:contribution-closure}), the combined output must carry an origin that downstream interfaces can use in the same way. The runtime therefore keeps the longest common prefix (LCP) of the participating paths and appends the new output's token:
\[
 P_{\mathrm{out},j}
   = \operatorname{LCP}\{P_e : e \in E_B\}\mathbin{\|}[t_{i,j}],
\]
where $E_B$ contains the inputs in bucket $B$ and $t_{i,j}$ identifies output $j$ of the combined execution at interface $i$. In Figure~\ref{fig:control-causality}, the three input paths are $C_0{:}D_0{:}H^a_0$, $C_0{:}D_1{:}H^b_0$, and $C_0{:}K_0$. Their LCP is $C_0$, so the first output of $J$ carries the path $C_0{:}J_0$. The internal $D/H/K$ branch history has been removed, while the shared origin $C_0$ remains available to downstream interfaces. The bucket identifier $B$ must remain fixed while inputs arrive; the current LCP cannot serve as that identifier because a later result may shorten it.

\begin{figure*}[t]
  \centering
  \includegraphics[width=\textwidth]{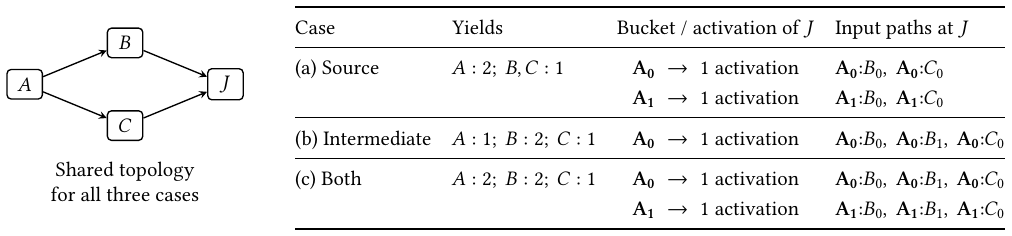}
  \caption{Prefix grouping in a diamond. Each row is one bucket, activating $J$ when readiness and ordering conditions hold. Bold prefixes identify source occurrences; yield counts are per activation. Paths use the text's abbreviated notation.}
  \label{fig:appendix-prefix-cases}
\end{figure*}

\paragraph{Repeated-yield examples.}
Figure~\ref{fig:appendix-prefix-cases} compares three output patterns on the same diamond $A\to B,C\to J$. All cases assume ordinary prefix joins with valid compiled anchors; count joins use a different scope.

\paragraph{Notation.}
$A$ is activated once, and yield counts are per activation. Paths such as $A_i{:}B_j$ omit activation identities and any shared outer prefix; the subscripts denote yield ordinals. Thus, $B_0$ under $A_0$ and $B_0$ under $A_1$ belong to different activations. Actual tokens retain full activation and event identities. The table lists contributions without prescribing their arrival or execution order.

\paragraph{Grouping repeated yields.}
The three cases differ in where repetition occurs. Each source yield $A_i$ identifies a separate bucket, whereas repeated intermediate yields accumulate in the bucket for one source occurrence. When both levels repeat, each $A_i$ still creates one bucket that collects all intermediate contributions. Thus, $J$ executes once per source occurrence rather than once per contribution. Grouping is independent of arrival order, and a bucket may close after its own contribution subtree resolves without waiting for later source outputs or for the source generator to terminate.

\paragraph{Contribution closure.}
A lease remains open while its covered $B$ activation can produce more contributions, and the corresponding slot closes only after that contribution subtree resolves. Different source occurrences maintain independent slot closure. A bucket becomes ready only after every slot root has resolved and all required fields are present; closure with a missing field makes the bucket impossible.

\paragraph{Input aggregation and admission.}
Multiple contributions to one field form an outer list in admission order; a single contribution retains its value. An application list remains one value and is not recursively flattened. For deliveries $d$ and $d'$, the admission rule is
\[
\operatorname{pred}(d)=d' \Longrightarrow
\operatorname{commit}(d') \prec \operatorname{admit}(d).
\]
Commitment of the predecessor contribution is distinct from completion of its activation. Deliveries produced by one yield for parallel routes share a predecessor rather than ordering one another. A skipped delivery, produced when a route supplies no required field, still completes its ordering obligation.

\subsection{Input Aggregation for Count-Based Activation (ABC)}
\label{appendix:count-contract}

This section specifies the selection and input-assembly rules for the request-scoped ABC joins introduced in Section~\ref{sec:count-join}. Each receiving interface uses one ABC bucket per graph request, shared across parser occurrences.

\paragraph{Selecting trigger events.}
For a trigger field with count $N$, the ABC bucket selects the first $N$ distinct event identities in reducer admission order. Delivering the same event through multiple input groups does not increase its count. If the count becomes known after contributions have arrived, selection follows their stored admission order. Selection freezes at $N$; a later event does not replace a selected one even if its path sorts earlier.

\paragraph{Matching companion fields.}
For each original input-group facet, the runtime unions its selected trigger events and deduplicates them by event identity to form an anchor set. Each non-trigger required field supplies one companion per anchor. A candidate must belong to the same facet and have an ancestor or descendant path under full token equality. The closest path-length distance wins; equally near distinct candidates are rejected as ambiguous. One ancestor contribution may serve multiple anchors, so trigger fields need not have equally long result lists and one companion value may appear multiple times in its field's output list.

\paragraph{Handling zero counts.}
A zero count selects no event for that trigger field. A facet with no selected trigger requires no companion and contributes no output group. A zero count does not activate an otherwise unvisited interface: incoming data events still drive bucket creation and readiness checks.

\paragraph{Ordering inputs and forming the output path.}
Selection precedes sorting. Paths compare their yield-ordinal vectors from outermost to innermost only when their complete node sequences match; paths of different shapes are incomparable, and stable topological sorting preserves admission order among ties. Each trigger field is sorted separately, companions follow the corresponding ordered anchors, and input-group facet boundaries are retained. Finally, the runtime folds the full token paths of all selected trigger and companion events and uses their structural longest common prefix as the output origin. Paths that are ordering-equivalent need not share a structural prefix.

\subsection{Distributed Execution and Request Lifecycle}
\label{appendix:control-execution}

Section~\ref{sec:control-flow} describes when an interface can execute. Interfaces may run asynchronously on different workers, and one interface may have multiple replicas. Within one graph call, related results sent to the same interface are routed to the same replica. For example, $h_0$, $h_1$, and $k_0$ in Figure~\ref{fig:control-causality} must reach the same replica of $J$ so that they can be collected in one bucket. The first delivery fixes this route, and subsequent deliveries to $J$ reuse it. A streaming interface sends each yielded result as it is produced and waits for the corresponding downstream delivery, providing backpressure while independent branches continue concurrently.

Omni-Flow tracks request-wide activity to determine when a graph call has finished. After the source has ended and no tracked interface execution or message delivery remains, the graph-call origin publishes the terminal result. A successful graph call ends with exactly one \texttt{SYSTEM\_FINISH}; a failed or cancelled call ends with an \texttt{ERROR} and does not emit \texttt{SYSTEM\_FINISH}. Heavier execution-state and resource cleanup can continue after this terminal result is published.

Graph calls in the same session are admitted in FIFO order, while calls in different sessions may proceed concurrently. A caller may also cancel earlier calls before admitting a new one. Cancellation uses an independent control path: it blocks new activations and outputs for the cancelled request and invokes available backend cancellation hooks. The cancellation can be acknowledged before remaining work and resources are fully cleaned up, without depending on progress through the application graph.

\section{Deployment Implementation}
\label{appendix:deployment}

\paragraph{Replica shapes and placement groups.}
A role declares its replica count, accelerator count per replica, node-span parameter, CPU budget, and placement layer. The launcher normalizes accelerator-role layer indices, then derives the placement-group (PG) shape from the lowest layer. Base-layer roles must agree on the accelerator count per bundle; their replica shapes determine the number of bundles. Higher layers are mapped onto devices already reserved by those bundles, subject to layout validation. Roles within a layer receive disjoint device slices, while different layers can reuse those slices. A fractional accelerator declaration in a higher layer selects physical devices; it does not establish a fractional GPU-memory or execution-time partition.

The inferred bundle count is distinct from the physical-node count. A placement strategy determines whether bundles share a node or spread across nodes; \texttt{STRICT\_SPREAD} requires distinct nodes. Thus, the resource configuration specifies a logical deployment layout rather than discovering an optimal cluster topology.

\paragraph{Reservation and device assignment.}
The PG reserves accelerator and CPU resources in Ray. Ordinary Python role actors launch with zero additional CPU allocation and no GPU allocation of their own, and are pinned to their assigned PG bundles. The launcher probes the devices assigned to each bundle and injects the appropriate device visibility and distributed-launch metadata into each role. This applies to base-layer and higher-layer actors alike: the distinction between layers determines the layout, not whether a role actor separately reserves GPUs. Resource reservation and process-level device visibility therefore have separate responsibilities.

\paragraph{Memory-manager scope.}
When memory management is enabled, the launcher creates a MemoryManager for each physical node used by a resource-group instance. Multiple bundles on the same identified node share one manager, whose device visibility covers their assigned devices. The launcher passes the corresponding manager endpoints and local device mappings to the roles. This connects physical placement to data-plane access; co-location alone neither deduplicates weights nor establishes KV compatibility. Those operations remain governed by the storage and adapter contracts described in the main text.

\paragraph{Configured replication.}
The role's \texttt{replicas} field sets the number of role replicas within a resource-group instance. The group's \texttt{min\_group\_count} sets the number of independently launched group instances, each with its own PG and associated managers and actors. These fields configure deployment-time replication; they do not by themselves provide load-driven online autoscaling or state-preserving migration. Such policies must also account for in-flight workflow state, session affinity, and reusable data.

\section{Model Integration Coverage}
\label{appendix:model-coverage}

Table~\ref{tab:model-coverage} summarizes 19 model integrations across nine task families. Each entry has model-specific implementation code, workflow definitions, and a CI configuration. Checkpoint variants within the same entry are grouped together. Integrations use the framework mechanisms appropriate to their computation and interaction patterns. This inventory describes implementation coverage; accuracy and performance results are reserved for Section~\ref{sec:evaluation}.

\begin{table}[t]
\caption{Model coverage: 19 integrations across nine task families. Variants are grouped within each entry.}
\label{tab:model-coverage}
\centering
\small
\setlength{\tabcolsep}{4pt}
\renewcommand{\arraystretch}{1.18}
\begin{tabular}{@{}p{0.29\columnwidth}p{0.66\columnwidth}@{}}
\toprule
\textbf{Task family} & \textbf{Model integrations} \\
\midrule
Language generation & DeepSeek-V2 \\
Speech recognition & LlamaMoE-ASR; Qwen3-ASR; Whisper (large-v3, large-v3-turbo) \\
Speech synthesis & Qwen3-TTS; Fish Speech S2-Pro \\
Vision-language understanding & Qwen3-VL \\
Multimodal workflows & LongCat-Next; LongCat-Lion; Qwen3-Omni \\
Streaming interaction & MiniCPM-o~4.5; PersonaPlex / Moshi \\
Image/video and joint generation & HunyuanImage-3; BAGEL; Lance (image and video checkpoints); MiniMax~H3 (joint audio-video) \\
Vision-language-action policies & OpenVLA (vanilla, OFT); SmolVLA \\
World models & Cosmos3 (Edge, Nano) \\
\bottomrule
\end{tabular}
\end{table}

The task families highlight breadth rather than mutually exclusive capabilities. For example, image-generation integrations can also include understanding paths. The OpenVLA entry includes autoregressive and one-forward action policies, while SmolVLA uses iterative action generation conditioned on a vision-language backbone. Cosmos3 includes generation and reasoning workflows. This diversity exercises a common framework without requiring all integrations to adopt the same scheduler or state-sharing pattern.

\section{Framework Runtime Overhead}
\label{appendix:overhead}

We use CPU microbenchmarks on the Intel Xeon Platinum 8480+ to measure control-flow scheduling and data-flow resource
management. Their timing units are one graph request and one task's resource-access cycle,
respectively. We additionally profile history-metadata management during
live model serving. These measurements exclude model computation and cannot
be summed into an end-to-end request overhead.

\paragraph{Control-flow scheduling.}
We use all 39 graph definitions in the repository's examples, preserving their
structure while replacing model computation with no-ops returning fixed
markers. No resource access or tensor transfer is performed. We measure pure control-flow overhead
from request entry through graph completion, covering dependency checks, event
propagation, task dispatch, and completion handling. We warm up each graph
20 times and measure
300 requests per graph.

\begin{figure}[t]
\centering
\includegraphics[width=\columnwidth]{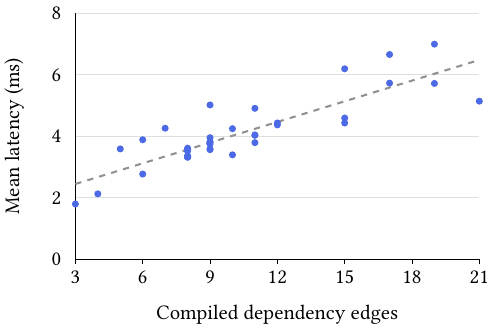}\par
\caption{Control-flow overhead.}
\label{fig:control-overhead}
\end{figure}

Figure~\ref{fig:control-overhead} plots mean latency against compiled
scheduling-dependency edge count, which ranges from 3 to 21; each point represents one graph, and the dashed gray line is a least-squares
linear fit.
Latency generally increases with edge count, but graph structure and
the exercised path also affect it. These results show that Omni-Flow supports
diverse model workflows while maintaining a mean pure control-flow overhead
of 1.79--7.00 ms across the 39 evaluated graph definitions.

\paragraph{Data-flow resource management.}
We measure a task's acquisition, metadata update, and release of 1--3 resident
resources, matching the range used by node calls in the examples. The resource count is the number
of independently named resources accessed by one task.
For example, accessing both \texttt{history\_info} and \texttt{kv\_cache}
counts as two resources. Timing
includes client--manager communication and management processing, but excludes
data movement, swapping, and resource destruction. Each resource starts with
one page; the update records a preallocated second page. Runs reset to the
same state and execute sequentially without contention, with 20 warm-ups and
300 measurements per configuration. Table~\ref{tab:resource-overhead} reports
phase means and directly timed totals: mean total latency is 3.72--10.98 ms.

\begin{table}[t]
\caption{Resident-resource management latency (ms). Total latency is measured
directly; P95 is its 95th percentile.}
\label{tab:resource-overhead}
\centering
\setlength{\tabcolsep}{3pt}
\begin{tabular}{@{}rrrrrr@{}}
\toprule
Resources & Acquire & Update & Release & Total & P95 \\
\midrule
1 & 1.654 & 0.669 & 1.396 & 3.720 & 4.979 \\
2 & 3.818 & 1.536 & 1.956 & 7.312 & 9.134 \\
3 & 6.400 & 2.117 & 2.462 & 10.980 & 13.908 \\
\bottomrule
\end{tabular}
\end{table}

\begin{figure}[t]
\centering
\includegraphics[width=\columnwidth]{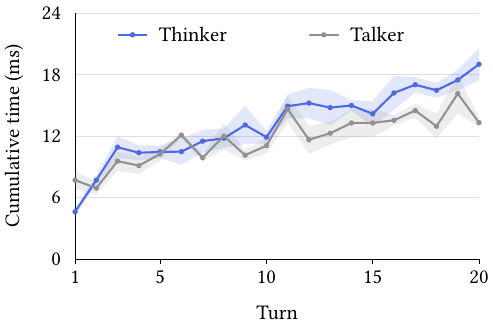}\par
\caption{History-metadata management overhead.}
\label{fig:history-metadata-overhead}
\end{figure}

\paragraph{Compute-flow history-metadata management.}
We measure Qwen3-Omni's history-metadata management overhead across
five 20-turn multimodal conversations, covering history-boundary retrieval and
validation, retention/truncation checks, and session-metadata updates after
appends. Timing includes communication for metadata queries but excludes
tensor writes/transfers, model and encoder computation, and KV prefix queries.
Figure~\ref{fig:history-metadata-overhead} reports the arithmetic mean of
per-turn cumulative management time across the five sessions, with shaded
bands indicating $\pm1$ sample standard deviation. Averaged over 20 turns,
Thinker and Talker take 13.16 ms and 11.72 ms, respectively. All five sessions use the same inputs and application settings and start
from the same initial state; streaming lengths and update counts may vary.
History maintenance reads the history-page list and checks page availability,
so growing histories increase the number of pages processed. 

\section{Workflow Definitions in Existing Frameworks}
\label{appendix:workflow}

The following examples illustrate how LangGraph, vLLM-Omni, and SGLang-Omni define and connect workflow stages.

\subsection*{LangGraph}
\label{appendix:langgraph}

\begin{lstlisting}[language=Python]
# Nodes = Python functions,
# Edges = add_edge / add_conditional_edges,
# State = TypedDict
import operator
from typing import TypedDict, Annotated, Any
from langgraph.graph import StateGraph, START, END

class S(TypedDict):
    prompt: str; image: Any; gen_audio: bool
    messages: Annotated[list, operator.add]
    streamed: Annotated[list, operator.add]
    audio: bytes

def preprocess(s):  return s
def img_enc(s):
    return {"img_emb": f"vit({s.get('image','')})"}
def embed(s):
    return {"txt_emb": f"emb({s['prompt'][:10]})"}
def thinker(s):
    return {
        "messages": [{"role":"assistant",
            "content":f"ans:{s['prompt'][:10]}"}],
        "streamed": ["tok1","tok2","tok3"]
    }
def talker(s):
    if not s.get("gen_audio"): return {}
    return {"audio":
        f"tts({s['messages'][-1]['content'][:10]})"
        .encode()}
def tools(s):
    return {"messages":
        [{"role":"tool","content":"result"}]}

g = StateGraph(S)
g.add_node("pre", preprocess)
g.add_node("img", img_enc)
g.add_node("emb", embed)
g.add_node("think", thinker)
g.add_node("tts", talker)
g.add_node("tools", tools)

g.add_edge(START, "pre")
g.add_edge("pre", "emb")
# Dynamic branching
g.add_conditional_edges("pre",
    lambda s: ["img","emb"]
        if s.get("image") else ["emb"],)
g.add_edge("img", "think")
g.add_edge("emb", "think")
g.add_edge("think", "tools")
g.add_conditional_edges("tools",
    lambda s: "think"
        if s["messages"][-1].get("tool_calls")
        else END,
    {"think": "think", END: END})
g.add_conditional_edges("think",
    lambda s: ["tts"]
        if s.get("gen_audio") else [END],)
g.add_edge("tts", END)
graph = g.compile()

async for e in graph.astream_events(
        INPUT, version="v2"):
    if (e["event"] == "on_chain_end"
            and "streamed" in
            e["data"].get("output",{})):
        print(e["data"]["output"]["streamed"])
\end{lstlisting}

\subsection*{vllm-omni}

\begin{lstlisting}[language=Python]
# Nodes = frozen dataclass,
# Edges = input_sources tuple,
# Bridge = function reference

P = PipelineConfig(model_type="m", stages=(
    StagePipelineConfig(
        stage_id=0, model_stage="thinker",
        execution_type=
            StageExecutionType.LLM_AR,
        input_sources=(),
        requires_multimodal_data=True,
        engine_output_type="text",
    ),
    StagePipelineConfig(
        stage_id=1, model_stage="talker",
        execution_type=
            StageExecutionType.LLM_GENERATION,
        input_sources=(0,),
        engine_output_type="audio",
        async_chunk_process_next_stage_input_func=
            "m.thinker2talker_chunk",
        sync_process_input_func=
            "m.thinker2talker_sync",
    ),
))

def thinker2talker_sync(src, **kw):
    return [OmniTokensPrompt(
        prompt_token_ids=[0],
        additional_information={
            "text": src[0].outputs[0].text})]
\end{lstlisting}

\subsection*{sglang-omni}

\begin{lstlisting}[language=Python]
# Nodes = JSON StageConfig + factory functions,
# Edges = next/route_fn/wait_for/stream_to

# pipeline.json
{
  "stages": [
    {"name":"pre","factory":"s.create_pre",
     "route_fn":"n.pre_next"},
    {"name":"img","factory":"s.create_img",
     "next":"agg"},
    {"name":"agg","factory":"s.create_agg",
     "wait_for":["pre","img"],"next":"think"},
    {"name":"think","factory":"s.create_think",
     "stream_to":["tts","dec"]},
    {"name":"tts","factory":"s.create_tts",
     "terminal":true},
    {"name":"dec","factory":"s.create_dec",
     "terminal":true}
  ]
}

# next_stage.py
def pre_next(rid, out):
    s = load_state(out)
    st = ["agg"]
    if s.has_image(): st.insert(0, "img")
    if s.has_audio(): st.insert(0, "aud")
    return st

# stages.py
def create_think(model_path):
    eng = create_sglang_ar_engine(...)
    def _stream(payload, token):
        return {"token_id": int(token)}
    return EngineExecutor(eng,
        stream_builder=_stream)
\end{lstlisting}
\section{AI-Assisted Development Disclosure}
\label{appendix:ai-coding}

OpenAI Codex and Anthropic Claude were used extensively during Omni-Flow's development for code generation , test generation, and bug diagnosis, as well as for editing the manuscript. The authors defined the system architecture, interfaces, and requirements; reviewed and revised the generated code; and conducted the experiments and validated the reported results. The authors take full responsibility for the implementation and the contents of this paper.

\end{document}